\documentclass[manuscript,screen,nonacm]{acmart}

\AtBeginDocument{%
  }

\begin{document}

\title{Viral Images: Identifying Reprintings within 1.5 Million Photographs in Chronicling America}


\author{Bruno Buccalon}
\affiliation{%
  \institution{History Department, Rice University}
  \city{Houston}
  \country{USA}}
\email{buccalon@rice.edu}

\author{Yueran Sun}
\affiliation{%
  \institution{Lab for Computing Cultural Heritage, University of Washington}
  \city{Seattle}
  \country{USA}}
\email{yuerans@uw.edu}

\author{Benjamin Charles Germain Lee}
\affiliation{%
  \institution{Information School, University of Washington}
  \city{Seattle}
  \country{USA}}
\email{bcgl@uw.edu}

\renewcommand{\shortauthors}{Buccalon, Sun \& Lee}

\begin{abstract}
Within the millions of digitized historic American newspapers in the Chronicling America initiative are tens of millions of photographs, illustrations, cartoons, and advertisements. Much of this visual culture is shared across newspaper titles and issues. Just as reprinted texts within these newspapers speak to the virality of textual content, so too does this reprinted visual culture speak to newspapers as sites of constant information circulation and exchange. In this paper, we introduce Viral Images, a project to identify reprintings within 1.5 million photographs in Chronicling America. For our analysis, we adopt the Newspaper Navigator dataset of extracted photographs from over 16 million pages in Chronicling America. We introduce an unsupervised method of identifying reprintings by leveraging contrastive language-image pretraining (CLIP) to embed these 1.5 million photographs and applying clustering to identify reprinted content. We detail our public interface, \url{https://viral-images.org}, which we designed in order to enable humanists to interactively browse and study these identified clusters. In addition, we analyze the identified clusters, uncovering a diversity of photographs and advertisements that have been circulated across different newspapers over time.
\end{abstract}



\maketitle

\section{Introduction}

Historic newspapers are troves of visual culture, regularly containing photographs, illustrations, editorial cartoons, infographics, maps, and other visual forms. This visual culture has far-ranging implications for scholarship within the humanities: political cartoons shed light on international relations and domestic policy; photographs reveal prominent national and local figures; maps demonstrate evolving borders and names. Significantly, this visual culture exists not only within the context of a newspaper page or title but also within the context of its \textit{circulation}: its transmission across different newspaper titles across time.

\begin{figure}
    \centering
    \includegraphics[width=1\linewidth]{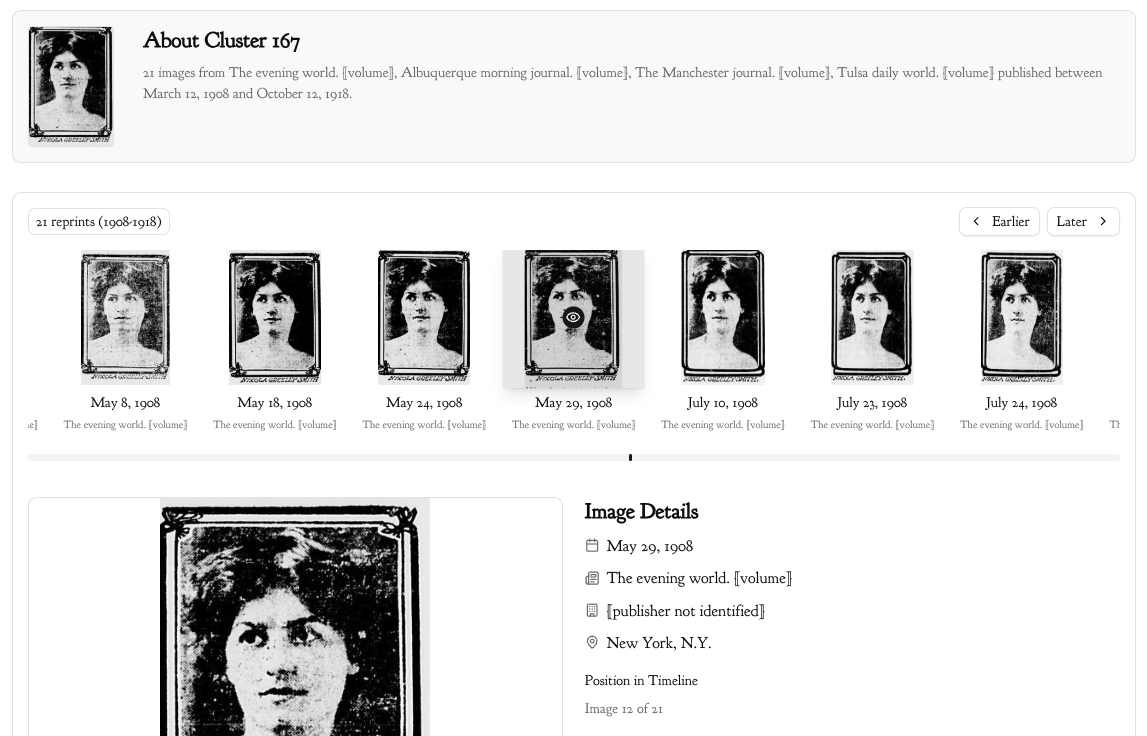}
    \caption{Cluster detail page of \url{https://viral-images.org} with image reprints ordered by publication date.}
    \label{fig:detailed-view}
\end{figure}

Collaborations such as the Viral Texts project have given much attention to developing methodologies for tracing the circulation of text passages throughout different newspaper titles, revealing network structures and patterns of distortion as text is used and re-used \cite{going_the_rounds}. But what about the visual culture? 

In this paper, we introduce Viral Images, a project to trace the circulation of images throughout historic newspaper pages. A circulated image may have different visual qualities between two digitized newspaper pages due to a series of factors, ranging from the materiality of the page (different inks, tears or stains on the page) to settings used when imaging from microfilm. To address this, we introduce an unsupervised method for clustering reprinted photographs among unsorted collections of images that is robust to distortions in the circulated image. To accomplish this, we rely on the Newspaper Navigator dataset of extracted visual culture from the Chronicling America newspaper collection \cite{news_nav_dataset}. In particular, we leverage the multimodal model CLIP \cite{radford2021learningtransferablevisualmodels} to embed all 1.5 million photographs from the dataset, and we apply DBScan  to identify clusters of reproduced photographs using these embeddings \cite{dbscan}.

Next, we analyze the clusters that we have identified using a mixed-methods approach. This analysis reveals patterns in the structural properties of clusters, the semantic categories of circulated images, and the ways photographs spread across newspapers and time. Lastly, to facilitate the analysis of these identified clusters of reprinted photographs for humanists, we publicly launch \url{https://viral-images.org}. With our public interface, we provide multiple affordances for browsing and analyzing these clusters, including faceted filtering across date range, newspaper title, and publisher. As shown in Figure \ref{fig:detailed-view}, we also provide affordances for the close reading of each cluster of reprinted photographs, enabling end-users to study image reprints alongside their metadata and to visualize a timeline of clustered images listed chronologically.

To facilitate reproducibility and re-use, we release all code for this project, from the clustering code to our containerized site, in an open-source repository: {\color{blue}{\url{https://github.com/L4CCH/viral-images}}}.

\subsection{Contributions}
In summary, our contributions are as follows:

\begin{enumerate}
    \item We introduce an unsupervised method for identifying reprinted photographs within historic newspapers, extensible across digitized archives writ large.
    \item We apply this method to historic photographs within the Newspaper Navigator dataset and identify clusters across 1.5 million photographs. We also introduce CLIP embeddings for all 1.5 million photographs in the dataset.
    \item We analyze and group the clusters according to their semantic properties.
    \item We publicly release a web application for browsing our identified clusters to facilitate humanistic analysis and close reading of the photographs. The website can be found at \url{https://viral-images.org}.

\end{enumerate}

\section{Related Work}

\subsection{Chronicling America \& Newspaper Navigator}

This project makes use of the Chronicling America database, containing over 23 million digitized historic newspapers published in the United States to date, along with the Newspaper Navigator dataset of extracted visual culture \cite{news_nav_dataset}. Created in 2020, the Newspaper Navigator dataset consists of extracted photographs, illustrations, comics, editorial cartoons, maps, headlines, and advertisements from 16 million pages available in Chronicling America at the time. The dataset was created using an object detection model that had been finetuned on crowdsourced annotations of visual content in Chronicling America produced by volunteers through the Library of Congress crowdsourcing initiative, Beyond Words. Currently, a version of the dataset is hosted via HuggingFace.\footnote{The current version of the dataset available on HuggingFace through the initiative BigLAM: BigScience Libraries, Archives and Museums. It contains only images published between 1850 and 1950: \url{https://huggingface.co/datasets/biglam/newspaper-navigator}.}
In this paper, we consider the 1.5 million photographs in the dataset. To learn more about the construction of the dataset, we refer the reader to \cite{news_nav_dataset}.

In 2020, the Library of Congress launched the Newspaper Navigator search application containing these 1.5 million photographs from the dataset \cite{news_nav_search_app}. In addition to supporting keyword search over the captions of the photographs (enabled via the Chronicling America OCR), the search application also enabled end-users to create and train ``AI navigators'' to search the collection visually. By iteratively defining positive and negative examples of what the end-user wanted to find and avoid, respectively, the end-user could iteratively steer the AI navigator toward content of visual interest. Using the search application led to the inevitable observation that many of the searches (both by keyword search over caption and by visual similarity with an AI navigator) yielded reprinted photographs. For example, the data archaeology of the Newspaper Navigator dataset explores four different reprintings of the same photograph of W.E.B. Du Bois appearing in Black newspapers in Chronicling America \cite{lee_compounded_2020}. Observing the abundance of these reprinted photographs, along with the results from the Viral Texts project (described in Section \ref{sec:virality}), was the initial motivation for this paper.

We note that because the photographs were extracted and classified by a machine learning model, the collection of photographs contains both false positives (for example, advertisements identified as photographs) as well as false negatives (photographs that are not included in the dataset). As described in later sections, many of our identified clusters actually contain advertisements with photographic elements.

\subsection{Virality in Historic Newspapers}\label{sec:virality}

Our work builds on a significant body of work exploring text reprintings and circulation within historic newspapers. The Viral Texts digital humanities project is the most direct inspiration for the work presented in this paper. Viral Texts charts reprintings of textual passages within \textit{Chronicling America} and other sources, including magazines \cite{blankenship, speculative_bib, smith_cordell_compuatational, detecting_viral, detecting_social, infectious_texts}. As noted by Cordell in \textit{Going the Rounds}, the reprinting of textual content in 19th-century American newspapers was a well-understood practice: ``As they combed through exchange papers (as well as magazines and books) in search of selections, nineteenth-century newspaper editors could spot which texts were being reprinted over and again by their peers. Editors often prefaced their reprintings of popular selections by telling readers a given text was going the rounds of the press or `going the rounds of the papers'' \cite{going_the_rounds}. Uncovering patterns of reprinting thus informs our understanding of this history of editorship, the formation of literary canon, and the information networks on which newspapers were built. Cordell proposes ``that newspaper selections were ancestors of modern `viral media,' embedded in an early platform of mass cultural production'' \cite{going_the_rounds}. Viral texts adopts approaches from natural language processing to identify these reprinting patterns, resulting in over 100,000 identified clusters. The project also offers invaluable analysis of the networks of reprinting and the significance of circulation for popular passages.

In this paper, we adopt an analogous approach to the visual culture within \textit{Chronicling America}. Our computer vision-based approach using CLIP \cite{radford2021learningtransferablevisualmodels} yields thousands of photograph clusters, and we demonstrate proof-of-concept analysis that similar modes of analysis can be applied to these visual clusters. We envision this work as a first step in a larger body of research to computationally analyze circulation among the millions of images found within historic newspapers.

This paper builds on other work surrounding the analysis of visual culture -- and visual features -- of  historic newspaper pages. For example, the Impresso project has built interfaces for browsing similar images, including image reprintings \cite{impresso-image} (the Impresso project has also explored patterns of text re-use in historic newspapers \cite{impresso-text}); Lee has analyzed the visual culture in the Ladino press, identifying clusters of reprinted advertisements and photographs \cite{ladino}; Lorang et al. have  utilized computer vision methods to identify poetry and other forms of content in historic newspaper pages \cite{ahr}.

\subsection{Circulation of Visual Culture}

Our project, Viral Images, is directly situated within circulation studies.
For example, Morina \& Bernstein trace the origins of internet memes through identifying patterns of circulation and alteration in born-digital images \cite{bernstein}. Smits and Ros utilize a computer vision API to trace the 940,000 online circulations of 26 famous photographs \cite{distant_raeading}. Gries applies digital methods to track the Shepard Fairey's \textit{Obama Hope} image \cite{gries}. Our project demonstrates a methodology to identify reprintings in an unsupervised manner, requiring no seed images.

\subsection{Computer Vision \& Cultural Heritage}

More broadly, our work builds on emerging research surrounding the application of computer vision models -- and more recently, multimodal models -- to digital collections from galleries, libraries, archives, and museums (GLAMs). The past decade of research within AI for GLAMs has demonstrated significant potential for the application of computer vision models to browsing collections, as evidenced by the capacity to search and analyze visual culture and even audiovisual content as far-ranging as newspapers \cite{news_nav_dataset, news_nav_search_app} to sitcoms \cite{distant_2023}. Just as ``distant reading'' has offered new epistemological approaches to literary studies, ``distant viewing'' -- a term coined by Arnold and Tilton -- sets a framework for how such AI-based approaches can enable visual analysis at a scale impossible for any single human to achieve \cite{distant_2023}. This turn toward computational analysis of visual features has been called the ``visual digital turn'' by Wevers and Smits \cite{visual_digital}. More recently, the advent of multimodal models -- models that learn jointly over images and text, embedding them into the same space -- such as contrastive language-image pre-training (CLIP) \cite{radford2021learningtransferablevisualmodels} has resulted in a ``multimodal turn'' \cite{multimodal_turn}, including a range of works that have explored the application of CLIP and other models to digital collections \cite{mahowald_lee, mapras, smits_JOHD, multimodal_turn, smits_kestemont, wevers_photos, digital_collections_explorer}.


\section{Methodology 1: Identifying Clusters of Reprintings}\label{sec:identifying}

\subsection{CLIP Embedding Generation}

We utilized the \texttt{clip-ViT-B-32} model to generate embeddings for all 1,568,530 photographs in the Newspaper Navigator dataset \cite{news_nav_dataset}. All computing was performed using an Amazon AWS \texttt{g4dn.12xlarge} instance with 48 Intel Cascade Lke vCPUs and 1 Nvidia T4 GPU.\footnote{We note that we generated CLIP embeddings using a locally-stored  version of the Newspaper Navigator dataset derived from when it was still hosted online by the Library of Congress, and this dataset is slightly larger than the HuggingFace dataset that we point to throughout this paper.} Processing embeddings for all 1.568 million images required approximately 8 hours of runtime.\footnote{Timing tests for 10,000 images averaged to 184 seconds, translating to 28,861 seconds, or just over 8.02 hours of runtime.} Given that the hourly cost of the \texttt{g4dn.12xlarge} is \$3.912 USD, this amounts to approximately \$31 USD for the total embedding pipeline.\footnote{We note that while our code utilized multiprocessing, we did not fully saturate CPU \& GPU usage, meaning that the overall compute -- and thus, compute cost -- could very well be lowered further.} All embeddings are available on HuggingFace at: {\color{blue}{\url{https://huggingface.co/datasets/L4CCH/viral-images}}}. 


We note that these CLIP embeddings can be utilized for a range of downstream tasks, including search \& recommendation, classification, and beyond.

\subsection{DBSCAN Clustering}

After generating embeddings, we applied the unsupervised clustering method DBSCAN (Density-Based Spatial Clustering of Applications with Noise) to group images in the embedding space \cite{dbscan}. We chose DBScan because it does not require the number of clusters to be specified in advance, an essential feature for our application. Moreover, DBScan is known to handle outliers well.

We utilized an AWS \texttt{c6a.48xlarge} EC2 instance with 192 vCPUs and 384 GiB of memory for clustering experiments. 
We ran DBScan using the Scikit-learn implementation \cite{scikit}; though we set \texttt{n\_jobs} = -1 with the goal of saturating CPU usage, we consistently found only half the cores were saturated (memory usage hovered consistently around 5GiB). Clustering all 1.56+ million images required just over 3 hours of compute, amounting to approximately \$22 USD spent on \texttt{c6a.48xlarge} instances (\texttt{c6a.48xlarge} instances cost \$7.344 per hour in the \texttt{us-east-2} region at the time of running the clustering). 

We utilized the default parameter setting of $min\_samples=5$ and selected our neighborhood radius $\epsilon = 2.4$ by inspecting the cluster outputs for a range of radii. Our qualitative assessment showed that smaller $\epsilon$ radii yielded smaller clusters due to false negatives, and larger $\epsilon$ radii merged clusters that did not belong together. We note that active future work includes quantitatively evaluating the optimal choice of $\epsilon$ and $min\_samples$, as well as different embedding choices, such as perceptual hashing and SigLIP \cite{zhai2023sigmoidlosslanguageimage}. 

Excluding noise, DBScan produced 1,910 non-noise clusters containing 20,753 images (each clustered image appears in precisely one cluster). In the next section, we provide quantitative analysis of identified clusters with $\epsilon = 2.4$. All clusters are made available in our repository: {\color{blue}{\url{https://github.com/L4CCH/viral-images}}}. 

\section{Methodology 2: Analyzing Viral Images}\label{sec:analyzing}

\subsection{Overview of Cluster Analysis}
In this  section, we detail exploratory analysis of identified clusters in order to examine their distribution, semantic content, and circulation patterns. 

\subsection{Cluster Size Distribution}
\begin{figure}[htbp]
    \centering
    \includegraphics[width=0.7\linewidth]{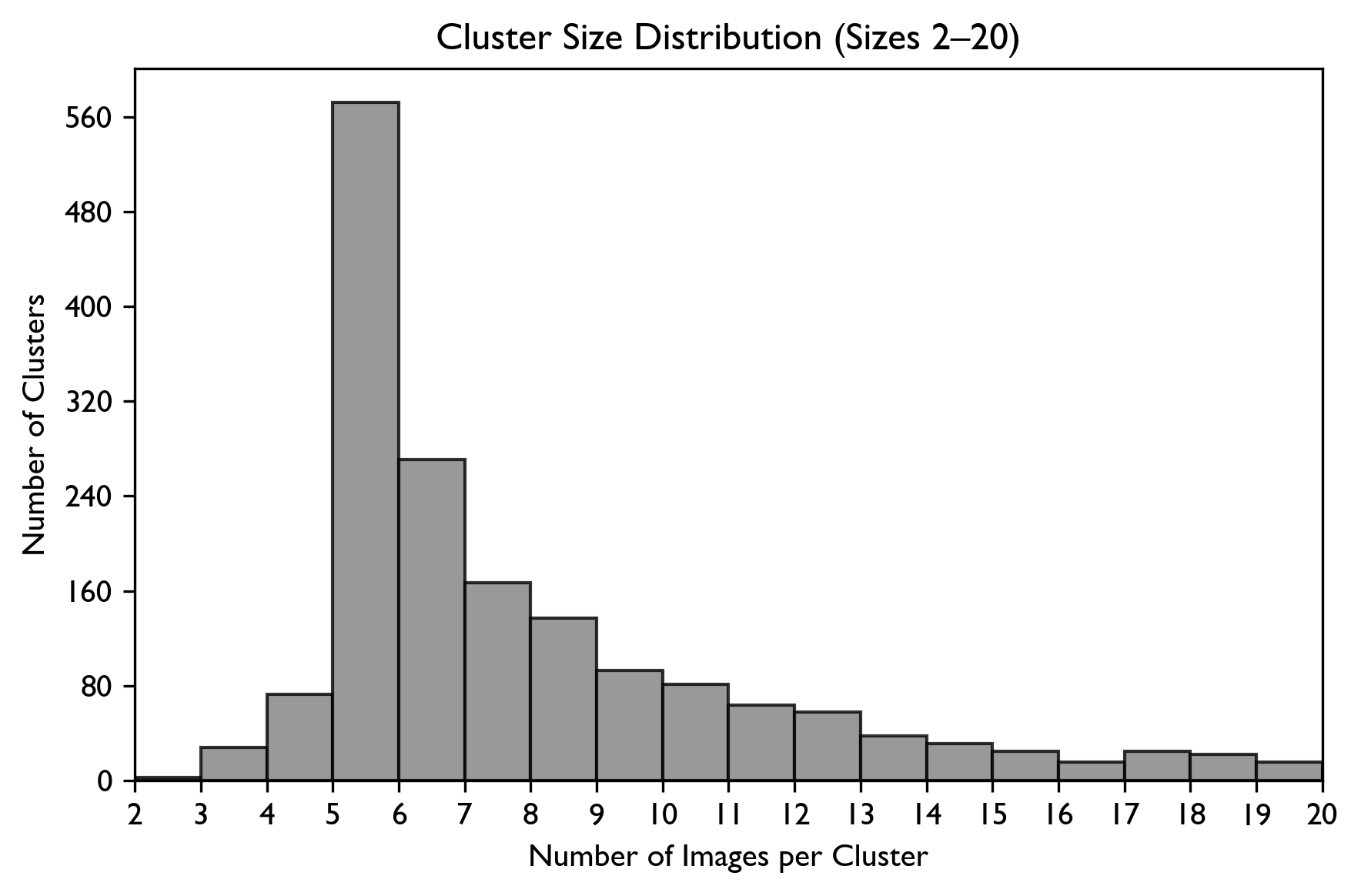}
    \caption{Distribution of cluster sizes between 2 and 20 images.}
    \label{fig:clustersize}
\end{figure}

In Figure \ref{fig:clustersize}, we display a histogram of cluster sizes, ranging from 2 to 20 images. This accounts for 1,731 clusters, or ~91\% of all clusters identified with DBScan. Of the remaining 179 clusters, 8 contain 100 images or more. The mode cluster size is likely a result of the DBScan setting of $min\_samples=5$; even though this mode is an artifact of our clustering algorithm, we are not concerned about our loss of clusters of four or fewer images because our main interest is in highly circulated images across different newspaper titles.

\subsection{Semantic Content of Clusters}
To characterize the visual content of the clusters, we applied a concept labeling approach using the \texttt{clip-ViT-B-32} model \cite{radford2021learningtransferablevisualmodels}. Because CLIP embeds both images and text within a shared representation space, textual descriptions can be used as a form of 0-shot classification of images. We defined a set of candidate text concept labels to be embedded based on our qualitative assessment of common categories through our inspection of our identified clusters: 
\begin{itemize}
\item "portrait"
\item "group of people" 
\item "person standing"
\item "building"
\item "landscape"
\item "marketing advertisement"
\item "transportation vehicle"
\item "animal"
\end{itemize}

To assess the most relevant label for each cluster, we averaged the embeddings of all images within a given cluster to produce a representative embedding. We then computed cosine distance between the labels and the representative embedding and identified the concept with the highest similarity score. To account for images lying outside of these categories, we added a category of "other," which we assessed using a two-component Gaussian mixture model. In particular, we separated clusters into high-confidence and low-confidence matches, yielding a threshold of 0.2485. However, qualitative inspection revealed that this threshold caused many clusters containing marketing advertisements to be labeled ambiguously; for this reason, we adopted a slightly more permissive threshold of 0.24. In Table \ref{tab:concept_summary}, we show the resulting cluster labels. Notably, the vast majority of circulation surrounds portrait shots.

\begin{table}[h]
\centering
\begin{tabular}{lrrrr}
\hline
\textbf{Concept label} 
& \textbf{No. clusters} 
& \textbf{\% of clusters} 
& \textbf{No. total images} 
& \textbf{\% of total images} \\
\hline
portrait & 1477 & 77.33 & 16406 & 79.05 \\
other & 248 & 12.99 & 2520 & 12.14 \\
marketing advertisement & 64 & 3.35 & 558 & 2.69 \\
building & 46 & 2.41 & 613 & 2.95 \\
person standing & 42 & 2.20 & 252 & 1.21 \\
group of people & 17 & 0.89 & 159 & 0.77 \\
transportation vehicle & 10 & 0.52 & 90 & 0.43 \\
animal & 3 & 0.16 & 34 & 0.16 \\
landscape & 3 & 0.16 & 121 & 0.58 \\
\hline
\end{tabular}
\caption{Summary of zero-shot concept labeling results across visual clusters.}
\label{tab:concept_summary}
\end{table}

\begin{figure}[htbp]
    \centering
    \includegraphics[width=0.8\linewidth]{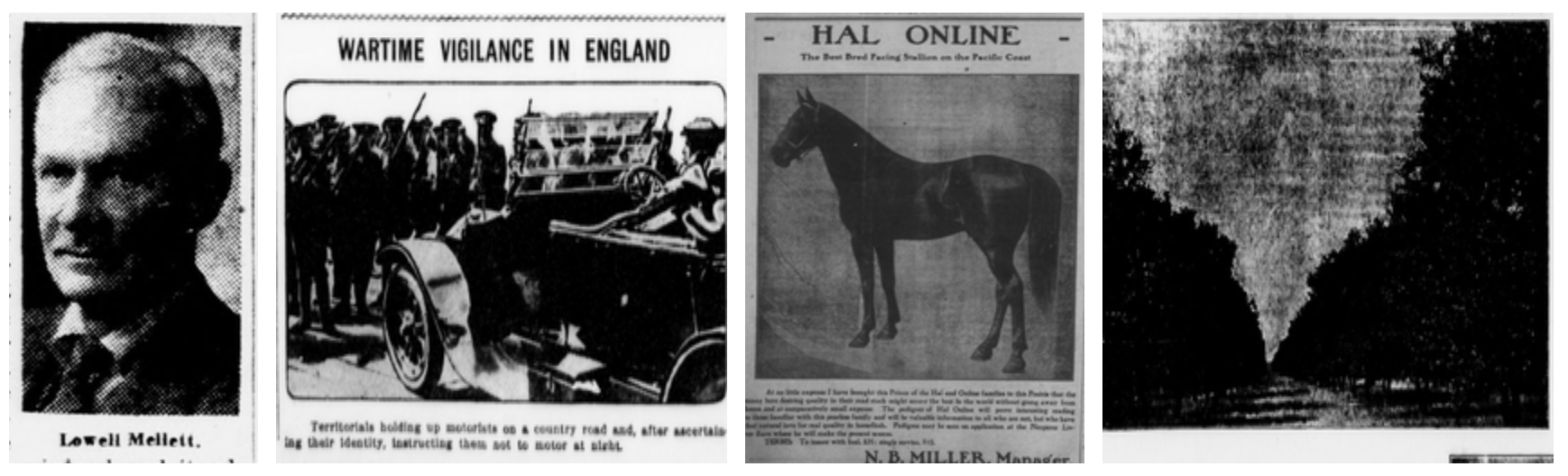}
    \caption{Examples of images categorized as "other."}
    \label{fig:otherex}
\end{figure}

Figure \ref{fig:otherex} shows several examples of images assigned to the "other" category. This includes images with poor print quality, heavy shadows over faces, or multiple relevant visual elements that elide clear categorization. Future work includes further investigating the nature of the photos in this category, as well as refining our full set of categories. Moreover, the lines between other categories are blurry: for example, some portrait shots are actually advertisements as well. Here, further multimodal analysis could leverage an image and its caption for classification.

\subsection{Cross-Newspaper and Temporal Circulation}

The circulation of images across not just different issues but different newspaper \textit{titles} is central to understanding the exchange of visual culture. Of the 1,910 clusters, 1,014 (53\%) appear in a single newspaper title, while 896 (47\%) span multiple titles.  In Table \ref{tab:cluster_title_temporal}, we show the top ten clusters, according to the number of newspaper titles containing at least one photo in the cluster. Nine out of these ten clusters fall into the portrait category. We also include a measure of \textit{temporal span}, computed as the number of days between the dates of publication of the earliest image and latest image in the cluster, revealing not just how much an image circulated but how long it circulated within \textit{Chronicling America}, our newspaper corpus of interest.

\begin{table}[htbp]
\centering
\begin{tabular}{rrrr}
\hline
\textbf{Cluster ID} & \textbf{No. Images} & \textbf{No. Unique Titles} & \textbf{Temporal Span (days)} \\
\hline
19 & 66 & 66 & 393 \\
14 & 77 & 64 & 495 \\
15 & 75 & 55 & 4950 \\
26 & 62 & 46 & 262 \\
44 & 45 & 45 & 48 \\
30 & 56 & 42 & 202 \\
54 & 40 & 40 & 82 \\
56 & 40 & 40 & 145 \\
35 & 51 & 38 & 643 \\
45 & 45 & 36 & 7665 \\
\hline
\end{tabular}
\caption{Top clusters ranked by the number of unique newspaper titles in which images appear, with their temporal spans.}
\label{tab:cluster_title_temporal}
\end{table}

\begin{figure}[htbp]
    \centering
    \includegraphics[width=0.7\linewidth]{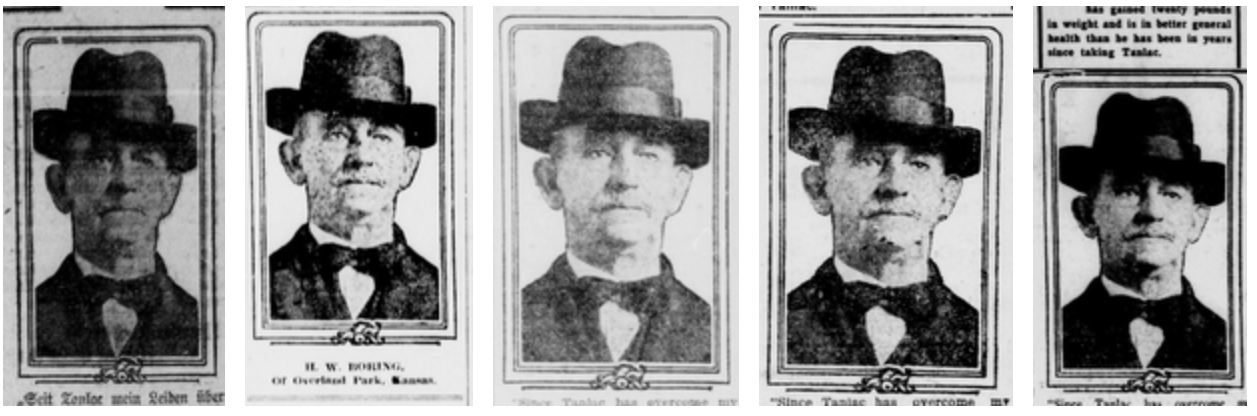}
    \caption{Sample images from Cluster 19: Portrait of the same individual appearing in 66 newspaper titles over 393 days.}
    \label{fig:cluster19}
\end{figure}

Figure \ref{fig:cluster19} shows example images from Cluster 19, which contains images from across the most newspaper titles. Each image is clearly the same portrait, with minor variations in print quality, resulting from reproduction across different newspapers. Notably, Cluster 15 spans 4,950 days ($\sim$13.5 years) and contains repeated portraits of former U.S. President William Taft, showing that images of prominent figures remained in circulation for many years. In contrast, Cluster 44 spans only 48 days. Inspection of these clusters suggests different modes of visual circulation: some images spread rapidly across multiple newspaper titles over a period of weeks, only to disappear from circulation, while others continued to circulate across the press network for more than a decade.

Although some clusters demonstrate clear patterns of re-use, inspecting clusters with the longest temporal spans (such as Cluster 45) reveals that different clusters of portraits have been merged together in some cases, yielding artificially large clusters. For example, some clusters group together images affected by ink saturation, and others contain multiple photos that look similar: for example, similar portrait formats or similar poses. Once again, future work entails refining our embedding and clustering approaches to reduce these errors. Moreover, as mentioned earlier in this section, directly leveraging captions could also improve results.
\section{Methodology 3: Building our Application}

While our methodology and analysis introduced in Sections \ref{sec:identifying} and \ref{sec:analyzing} provide new ways of studying the circulation of visual culture in historic newspaper pages, we also understand the importance of developing public interfaces that enable scholars to study the identified patterns of reprinting at different scales, ranging from close interrogation of individual reprintings within a cluster of circulation to viewing circulation trends across newspaper title, publisher, and publication date. In this section, we introduce our public website, \url{https://viral-images.org}, built with precisely these goals in mind.

\begin{figure}[htbp]
    \centering
    \includegraphics[width=0.8\linewidth]{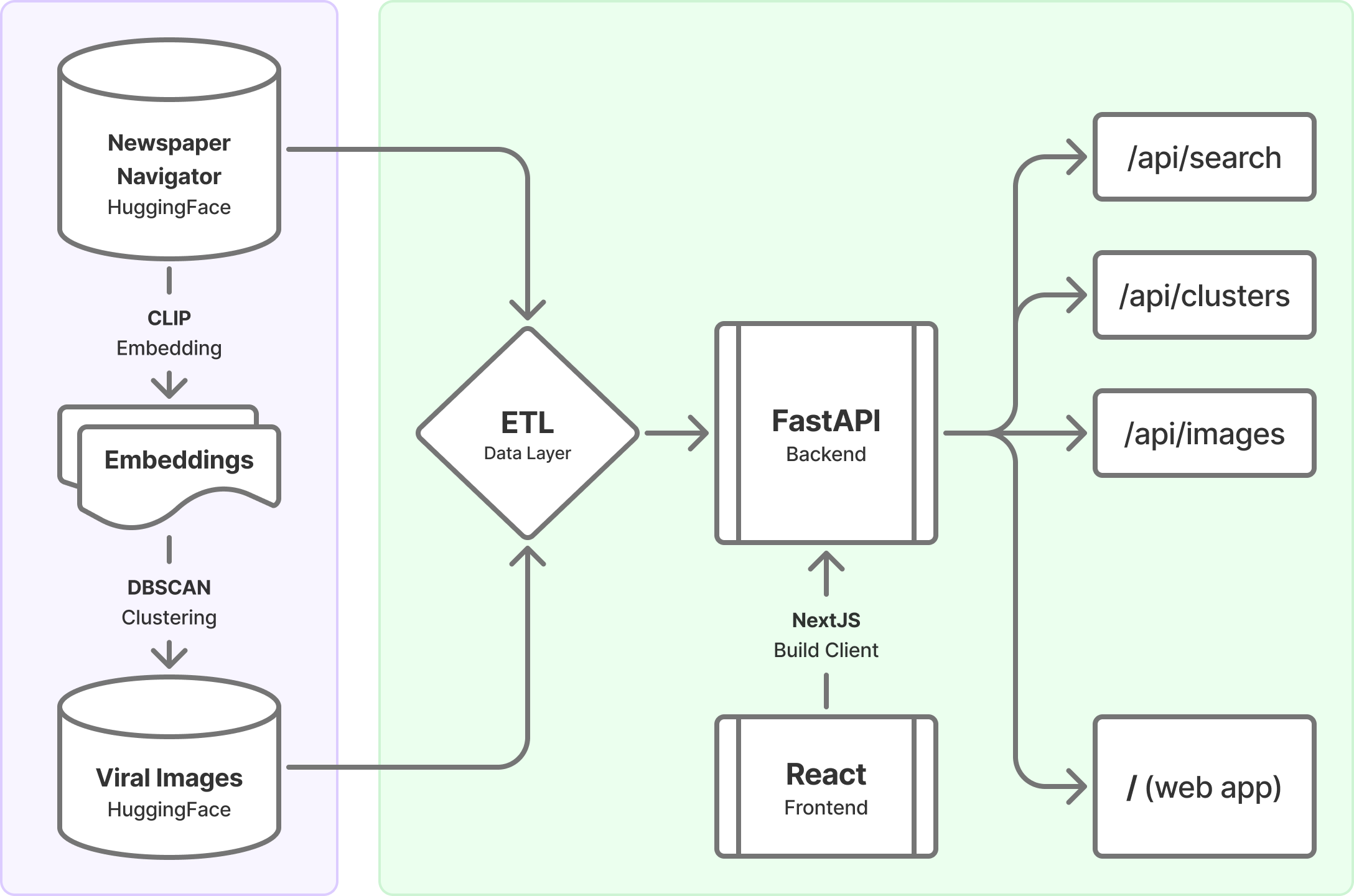}
    \caption{Diagram of the clustering process (purple, left side) and the web application architecture (green, right side).}
    \label{fig:backend_architecture}
\end{figure}

\subsection{System Architecture}

We had two main inspirations for the Viral Images web application: first, the Viral Texts website, in which each cluster is made available as a standalone page, enabling users to analyze and to cite specific clusters \cite{going_the_rounds}; and second, the Impresso web application, particularly in its affordances for displaying similar images in newspaper collections \cite{impresso-text}. We opted for a system architecture based on a FastAPI backend due to its easy integration with Python-based tools for machine learning. Our frontend client relies on the React and NextJS frameworks, with a user interface based on shadcn/ui components, resulting in a responsive Single Page Application that works well on mobile and desktop displays. Our architecture uses a multistage Dockerfile for preprocessing the dataset and exporting the frontend client at build time, so that both frontend and backend can be deployed as a single FastAPI application on AWS AppRunner. The web application architecture is represented in Figure \ref{fig:backend_architecture}.

\subsection{Affordances}

\begin{figure}
    \centering
    \includegraphics[width=0.9\linewidth]{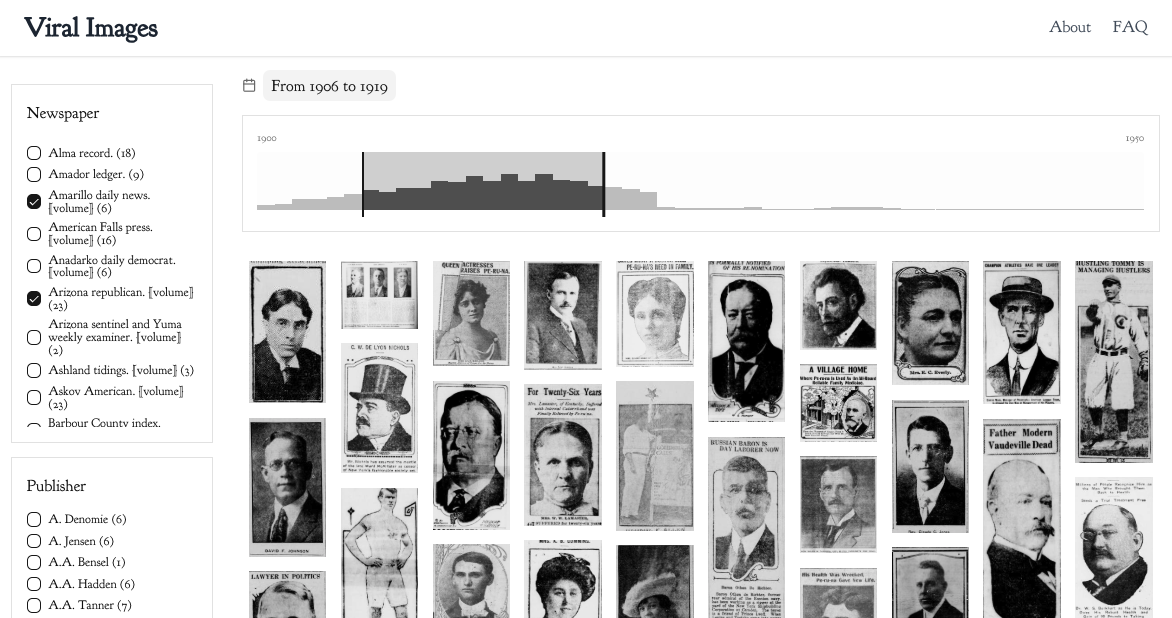}
    \caption{Gallery page with search filters by publication date, newspaper title, and publisher.}
    \label{fig:gallery-view}
\end{figure}

Our public interface provides multiple affordances for browsing and analyzing clusters. The initial page displays a gallery with search facets, so that users can filter the results by publication date, newspaper title, and publisher. Filters are based on image metadata available in the Newspaper Navigator dataset, so that each cluster contains the metadata of all its individual images. The gallery also displays a timeline plotting a histogram of cluster counts by publication date, allowing users to filter results by selecting a date range using drag-and-drop sliders. The gallery page with active search filters is represented in Figure \ref{fig:gallery-view}.

Once a cluster is selected, the user is redirected to a detailed page. Following the display of basic information about the cluster, the detail page presents a horizontal list of image reprints, ordered chronologically by publication date. Each reprint has a thumbnail along with the publication date and newspaper title. The first image in the cluster is selected by default so that users can see a larger version of the digitized image and their metadata as they scroll down the page. By selecting an image within the reprint timeline, the image viewer updates and displays the available metadata for that image. These affordances, that allow for the close reading of each cluster, are represented in Figure \ref{fig:detailed-view}.

\section{Conclusion}

In this paper, we have introduced our Viral Images, a project to identify image reprintings within historic newspaper pages. We have motivated our work with Viral Images within the context of computer vision, digital humanities, and circulation studies. Using the Newspaper Navigator dataset of extracted visual content from 16 million pages of digitized historic American newspapers in Chronicling America, we have introduced three central contributions. First, adopting 1.5 million photographs in the Newspaper Navigator dataset, we have demonstrated our unsupervised method of identifying reprintings. Second, we have provided analysis of the identified clusters of reprintings, revealing that most clusters are small and highly right-skewed in size, that portrait photographs dominate the semantic content of reused images, and that visual circulation occurs in multiple temporal patterns, including images that spread rapidly across many newspaper titles within weeks and others that remain in circulation for more than a decade. Third, we have introduced our Viral Images search application at: \url{https://viral-images.org}, which enables scholars and members of the public to interactively browse clusters of image reprintings and analyze them according to newspaper title, publisher, and date of publication. 

\subsection{Future Work}

We now turn to future directions for each of the components of the project. Surrounding the clustering method itself, we have three directions of ongoing and future work. First, we aim to more thoroughly evaluate clustering performance in a quantitative manner. Second, we will use these evaluation benchmarks to further finetune clustering hyperparameters and test clustering methods beyond DBScan. Third, we will test additional embedding models, including both visual models and multimodal models, such as vision-language models (which may benefit by identifying reprinted captions in the images). In addition, future work includes investigating the large number of images that were not assigned to any cluster. While some of these images may represent photographs that were never reprinted, others may correspond to reprints that are difficult to detect.

Regarding our website \url{https://viral-images.org}, we plan to incorporate additional functionality. For the front-end, we will allow users to reorder the gallery results by cluster size, such as image count or number of unique newspapers and publishers. We also hope to support the display of image reprints in their original publication context. By implementing IIIF Content State annotations \cite{appleby_iiif_2022}, which must rely on the International Image Interoperability Framework (IIIF) infrastructure maintained by the Library of Congress, our users will be able to see image reprints as bounding boxes displayed over the original newspaper page, while also having access to the entire newspaper edition. We will also implement CLIP-based search for text and image similarity, supporting a similar set functionality made available by the Digital Collections Explorer \cite{digital_collections_explorer}, such as natural language search and file upload for image similarity search. We also plan to conduct user-centered analysis with print scholars to understand what affordances could be improved with our website.

Within the digital humanities, we are also excited to work with print scholars to further contextualize and analyze reprinting patterns identified in historic newspaper pages. For example, we can analyze captions of circulated images to find  patterns and changes in description. Lastly, we can study how the circulation of visual culture correlates with the circulation of text by leveraging the Viral Texts findings. 

We encourage readers to visit \url{https://viral-images.org} and contact us with feedback and questions.

\section{Reproducibility Statement}

To ensure reproducibility and public availability, we have released all our code for Viral Images in an open-source codebase available at: {\color{blue}{\url{https://github.com/L4CCH/viral-images}}}.
This includes our clustering pipeline as well as the full Viral Images search. The full Newspaper Navigator Dataset can be found at: {\color{blue}{\url{https://huggingface.co/datasets/biglam/newspaper-navigator}}}. We note that we used Cursor, Gemini, and Claude while developing our codebase, as well as to assist in data analysis.


\bibliographystyle{ACM-Reference-Format}
\bibliography{bibliography}

\end{document}